\documentclass[twocolumn,prb,color,psfig,showkeys]{revtex4}
\usepackage{graphicx}
\usepackage{epsfig}
\usepackage{amsmath}
\usepackage{amssymb}

\newcommand{\eps}{\ensuremath{\varepsilon}}
\newcommand{\s}{\ensuremath{\sigma}}
\newcommand{\w}{\ensuremath{\omega}}
\newcommand{\p}[1]{#1^{\prime}}
\newcommand{\eq}[1]{(\ref{#1})}
\newcommand{\LSCO}{La$_{2-x}$Sr$_{x}$CuO$_{4}\,$}

\newcommand{\LSMO}{La$_{1-x}$Sr$_{x}$MnO$_{3}\,$}

\newcommand{\mb}[1]{\mathbf{#1}}

\begin{document}

\title{Infrared optical response of strongly correlated cuprates: \\ the effects
 of topological phase separation.}

\author{A.S. Moskvin}
\author{E.V. Zenkov}
\email{eugene.zenkov@usu.ru} \affiliation{Ural State University, 620083
Ekaterinburg, Russia}


\begin{abstract}

 Cuprates are believed to be unconventional systems which are
unstable with regard to a self-trapping  of the low-energy charge transfer
excitons with a nucleation of electron-hole (EH) droplets being actually the
system of coupled electron CuO$_{4}^{7-}$ and hole CuO$_{4}^{5-}$ centers
having been glued in lattice due to a strong electron-lattice polarization
effects. The hole/electron doping into parent cuprate is likely to be a driving
force for a growth of primary EH droplets with a gradual stabilization of  a
single, or multi-center skyrmion-like EH Bose liquid collective mode, followed
by a first order phase transition from parent insulating state into
unconventional topological EH Bose liquid phase (EHBL). Nanoscopic electron
inhomogeneity is believed to be inherent property of doped cuprates throughout
the phase diagram beginning from EH droplets in insulating parent system and
ending by a topological phase separation in EHBL phase.

We examine the effects of electron inhomogeneity accompanying such a phase
separation on IR optical conductivity. A simple model of metal-insulator
composite and effective medium theory has been used to describe the static
phase separation effects. The low-frequency dynamics of topological EHBL phase
in a random potential  in underdoped regime has been discussed in a
quasiparticle approximation in frames of the memory function formalism.

The effects of static and dynamic nanoscopic phase separation are believed to
describe the main peculiarities of the optical response of doped cuprates in a
wide spectral range.

\end{abstract}

\maketitle

\section{Introduction}

The nature of the optical response of strongly correlated cuprates is presently
still a matter of great controversy.
 The unconventional behavior of
cuprates strongly differs from that of  ordinary metals and merely resembles
that of a doped semiconductor. Moreover, the history of high T$_c$'s itself
shows that we deal with a transformation of particularly insulating state in
which the electron correlations govern the essential physics. New principles
must be developed to treat  these systems with its non-Fermi-liquid behavior.
First of all we have to change the current paradigm of the metal-to-insulator
(MI) transition to that of an insulator-to-metal (IM) phase transition. These
two approaches imply essentially different starting points: the former starts
from a rather simple metal scenario with inclusion of correlation effects,
while the latter does from strongly correlated atomic-like scenario with the
inclusion of a charge transfer.

The copper oxides start out life as insulators in contrast with BCS
superconductors being conventional metals. It is impossible to understand the
behavior of the doped cuprates and, in particular, the origin of HTSC unless
the nature of the doped insulating state is incorporated into the theory. In
particular, the Fermi liquid theory of the normal state and the BCS theory of
the superconducting state, which are so successful for conventional metals,
were not designed for doped insulators,  and they do not apply to the HTSC.
 Consequently it is necessary to develop a new mechanism and
many-body theory of HTSC.

The problem of a doped insulator is sure much more complicated than it is
implied in   oversimplified approaches such as an effective $t$-$J$-model.
First, we deal with effects of electron inhomogeneity and phase separation.
These  are currently considered to be inherent property of lightly doped or
so-called under-doped cuprates. However, recent scanning electron micrography
measurements clearly evidence the inherent electron inhomogeneity in optimally
doped  Bi$_2$Sr$_2$CaCu$_2$O$_8$. \cite{Pan,lang}

The emergence of electron inhomogeneity leads to a strong complication of the
optical response and its analysis due to a number of specific effects typical
for optically inhomogeneous systems. The discussion and modelling of such
effects would be the main goal of the paper. Its structure is organized as
follows. In Sec.II we discuss the origin and evolution of electron
inhomogeneity in doped cuprates. In Sec.III the effective medium theory is
applied to describe the optical response of "quenched" electronic
inhomogeneity. The effects of the dynamics of multi-center topological defect
in a random potential are addressed in Sec.IV.

\section{Condensation of CT excitons and phase separation in doped cuprates}
In  case of cuprates we deal with systems which conventional ground state seems
 to be unstable with regard the transformation into a new phase state with
a variety of unusual properties. It is worth noting the text-book example of
BaBiO$_3$ system where we unexpectedly deal with the disproportionated
Ba$^{3+}$+ Ba$^{5+}$ ground state instead of the conventional lattice of
Ba$^{4+}$ cations. The bismuthate situation can be viewed as a result of
condensation of charge transfer (CT) excitons, in other words, the  spontaneous
generation of self-trapped CT excitons in the ground state with a proper
transformation of lattice parameters. These problems are closely related with
the hidden multistability intrinsic to each solid. \cite{Toyozawa} If the
ground state of a solid is pseudo-degenerate, being composed of true and false
ground states with each structural and electronic orders different from others,
one might call it multi-stable.

In our view, the physics of the  doped cuprates, including  superconductivity,
is driven by a self-trapping of the CT excitons.\cite{Shluger,Vikhnin} Such
excitons are the result of self-consistent charge transfer and lattice
distortion with appearance of the ``negative-$U$'' effect. In contrast with
BaBiO$_3$ system where we deal with a  spontaneous generation of self-trapped
CT excitons (STE) in the ground state,   the parent insulating cuprates are
believed to be near excitonic instability when
 the self-trapped CT
excitons  form the candidate relaxed excited states to struggle with the
conventional ground state. \cite{Toyozawa} In other words, the lattice relaxed
CT excited state should be treated on an equal footing with the ground state.
If the interaction between STE were attractive and so large that the cohesive
energy $W_1$ per one STE exceeds the energy $E_R$ of one STE, the STE's and/or
its clusters will be spontaneously generated everywhere without any optical
excitation, and be condensed to form a new electronic state on a new lattice
structure. \cite{Toyozawa}

\subsection{Electron-hole droplets in cuprates: pseudo-impurity and phase separation regimes}

 Cuprates are believed to be unconventional systems which are
unstable with regard to a self-trapping  of the low-energy charge transfer
excitons \cite{Ng,Moskvin,Moskvin1} with a nucleation of EH droplets being
actually the system of coupled electron CuO$_{4}^{7-}$ and hole CuO$_{4}^{5-}$
centers having been glued in lattice due to a strong electron-lattice
polarization effects. Phase transition to novel hypothetically metallic state
could be realized due to a
  mechanism familiar to semiconductors with filled bands such as Ge and Si where
  given  certain conditions one observes
   a formation of metallic EH-liquid as a result of the exciton decay.
   \cite{Rice}
 However, the  system of strongly correlated electron CuO$_{4}^{7-}$ and hole
CuO$_{4}^{5-}$ centers  appears to be equivalent to an electron-hole
Bose-liquid (EHBL) in contrast with the electron-hole  Fermi-liquid in
conventional semiconductors. A simple model description of such a liquid
implies
 a system of local singlet bosons with a charge of $q=2e$ moving in a lattice
  formed by hole centers.\cite{1}
Local boson in our scenario represents the electron counterpart of Zhang-Rice
singlet, or two-electron configuration $b_{1g}^{2}{}^{1}A_{1g}$.
  Naturally, that conventional electron CuO$_{4}^{7-}$ center represents a
  relaxed state of a composite system: "hole CuO$_{4}^{5-}$ center plus local
  singlet boson", while the "non-retarded" scenario of a novel phase
is assumed to incorporate the unconventional states of electron CuO$_{4}^{7-}$
center up to its orbital degeneracy.

{\it Homogeneous nucleation} implies the spontaneous formation of EH droplets
due to the thermodynamic fluctuations in exciton gas. Generally speaking, such
a state with a nonzero volume fraction of EH droplets and the spontaneous
breaking of translational symmetry can be stable in nominally pure insulating
crystal. However, the level of intrinsic non-stoihiometry in oxides is
significant (one charged defect every 100-1000 molecular units is common). The
charged defect produces random electric field, which can be very large (up to
$10^8$ Vcm$^{-1}$) thus promoting the condensation of CT excitons and the
nucleation of EH droplets.
 Deviation from the neutrality of the CuO$_2$ layers implies the
existence of additional electron, or hole centers that can be the natural
 centers for the {\it inhomogeneous nucleation} of the EH droplets.
Such droplets are believed to provide the more effective screening of the
electrostatic repulsion for additional electron/hole centers, than the parent
insulating phase. As a result, the electron/hole injection to the insulating
cuprate due to a nonisovalent substitution as in La$_{2-x}$Sr$_x$CuO$_{4}$,
Nd$_{2-x}$Ce$_x$CuO$_{4}$, or change in oxygen stoihiometry as in
YBa$_2$Cu$_3$O$_{6+x}$, La$_{2}$CuO$_{4-\delta}$,
La$_{2}$Cu$_{1-x}$Li$_x$O$_{4}$, or field-effect is believed to shift the phase
equilibrium from the insulating state to the unconventional electron-hole Bose
liquid, or in other words  induce the insulator-to-EHBL phase transition.
%
%

The optimal way to the nucleation of EH droplets  in parent system like
La$_2$CuO$_4$, YBa$_2$Cu$_3$O$_6$ is to create charge inhomogeneity by
nonisovalent chemical substitution in CuO$_2$ planes or in ``out-of-plane
stuff'', including interstitial atoms and vacancies. This process results in an
increase of the energy of the parent phase and creates proper conditions for
its competing with others phases capable to provide an effective screening of
the charge inhomogeneity potential. The strongly degenerate system of electron
and hole centers in EH droplet is one of the most preferable ones for this
purpose. At the beginning (nucleation regime) an EH droplet nucleates as a
nanoscopic cluster composed of several number of neighboring electron and hole
centers pinned in the CuO$_2$ plane by disorder potential. Of course, the
conditions for such a nucleation  are delicately determined by the crystal and
chemical surroundings of the CuO$_2$ planes (``out-of-plane stuff''). Both the
electron and hole centers at the atomic-molecular level and the EH droplets at
the nanoscopic scale level are the elementary building blocks for an
unconventional electron-crystalline structure of the cuprates with the
nonisovalent substitution. EH droplet is a nanoscopic center for a pinning of
the various quasi-local hybrid charge-spin-vibronic modes with a definite
relaxation time and spatially non-uniform DOS distribution or even internal
micro-phase separation.

An occurrence of the EH droplets as the domains of the novel phase  manifests
itself in various properties of the cuprates with a non-isovalent substitution
even at the low content of the  nucleation centers. On the one hand, main
features of this pseudo-impurity regime are determined  by the intrinsic
properties of the EH droplets with expected delocalization and metallic-like
behavior at $T > T_{CO}$, localization with charge ordering at $T < T_{CO}$,
and other phase transformations up to a quasi-local bose-condensation at $T <
T_{BS}$ with an appearance of the superconducting order
parameter.\cite{PhysicaB} On the other hand, the real properties of the system
will be determined  by the peculiar geometrical factors as a relative volume
fraction of  the EH droplets, their average size and appropriate dispersion,
and, possibly, by their form and/or texture. Numerous examples of the
unconventional behavior of the cuprates in the pseudo-impurity regime could be
easily explained with taking into account the inter-phase boundary effects
(coercitivity, the mobility threshold, oscillations, relaxation etc.) and
corresponding characteristic quantities. In this connection one should
emphasize an occurrence of the characteristic temperatures T$_b$ for the start
of  the boundaries motion, or for the loss of the stability of  either
geometrical droplet configuration. In general, such phenomena should accompany
any partial phase transition for a phase-separated system, that could result in
an appearance of  a peculiar doublet structure of the temperature anomalies.

%
%

EH droplets  can manifest itself remarkably  in various properties of the
cuprates even  at  small volume fraction, or in a ``pseudo-impurity regime''.
Main features of this ``pseudo-impurity regime'' are determined  both by the
intrinsic properties of the EH droplets  and  by the peculiar geometrical
factors as a relative volume fraction of droplets, their average size and
appropriate dispersion, and, possibly, by their form and/or texture. Numerous
examples of the unconventional behavior of the cuprates in the pseudo-impurity
regime could be easily explained with taking into account the inter-phase
boundary effects (coercitivity, the mobility threshold, oscillations,
relaxation etc.) and corresponding characteristic quantities.

Under increasing doping the ``pseudo-impurity regime'' with a relatively small
volume fraction of EH droplets can gradually transform into a micro-
(electronic) or macro- (chemical) ``phase-separation regime'' with a sizeable
volume fraction of EH droplets.

The above viewpoint on the phase separation (PS) phenomena in cuprates is, in
general, compatible with the pioneer ideas by Emery and Kivelson
\cite{Emery2,EK,Emery3} and some other model approaches. So on the basis of the
neutron scattering data Egami \cite{Egami}  conjectured an appearance of a
nano-scale heterogeneous structure which is composed of a plenty mobile-carrier
existing region of metallic conductivity and semi-localized scarce carrier
region with antiferromagnetic spin ordering.

 Phase separation is now widely
discussed as an important phenomenon accompanying the high-T$_c$
superconductivity. The pseudo-impurity regime can be termed as a nanoscopic
phase separation.
Macroscopic, or chemical phase separation with the $x$-rays and neutrons
detectable domains of a new phase exactly takes place in super-oxygenated
La$_2$CuO$_{4+\delta}$ \cite{214d}, due to the enhanced mobility of negatively
charged oxygens which can compensate charge unbalance introduced by the hole
separation.  A tendency towards such a phase separation has also been
repeatedly observed in other cuprates.

At once, the well developed inhomogeneity and phase separation are inconsistent
with the excellent   coherence of the lattice demonstrated by the sharpness of
many of the Bragg peaks (for example, coherence length over 1000$\AA$  in
La$_{2-x}$M$_x$CuO$_4$ \cite{diffuse}). However, an appearance of the clearly
seen nonphonon  X-ray diffuse scattering with unconventional temperature and
${\bf q}$-dependence \cite{diffuse}  evidences an occurrence of the local or
quasi-local  (nanoscale) static and/or dynamic charge inhomogeneity as a
possible result of  the electronic phase separation. Lack of significant
elastic neutron diffuse scattering \cite{diffuse}  is compatible with the
mainly electronic nature of the inhomogeneity in doped cuprates.

%
%

\subsection{Topological phase separation in doped cuprates}
The phase separation and electron inhomogeneity are currently considered to be
inherent property of lightly doped or so-called under-doped cuprates. Our
scenario implies the existence of a critical doping level for the first-order
phase transition from parent insulating phase to EHBL phase. EHBL represents
the phase state with a condensation of CT excitons. Without any doping such a
phase is equivalent to a half-filled hard-core Bose-Hubbard (hc-BH) system.
Does new phase is homogeneous or not? We argue that EHBL phase is not only
inherently inhomogeneous but this inhomogeneity governs the low-energy physics
of doped cuprates. Indeed, the model of the EH Bose-liquid implies that  the
doped cuprates are proved to be in the universality
   class of the pseudo-spin 2D systems which description incorporates the
   inhomogeneity centers like
    topological defects to be a natural element  of essential physics.
    In addition, it should be noted that the self-trapping of CT excitons with nonzero electric dipole moment
     is characterized  by the nucleation of vortex-like configurations providing the minimum of the
    dipole-dipole coupling. In other words, the nucleation of the EH Bose
    liquid implies the formation of different topological textures.
    However, below we concentrate on the deviation from the half-filling to be
    seemingly the main driving force of the topological inhomogeneity in doped cuprates.

%
%

 The hc-BH-model  (quantum lattice Bose gas) has a long history and has been
suggested initially for conventional superconductors \cite{Schafroth}, quantum
crystals such as $^{4}He$ where superfluidity coexists with crystalline order.
\cite{Matsuda,Fisher} Afterwards, the Bose-Hubbard  model has been studied as a
model of the superconductor-insulator transition in materials with  local
bosons, bipolarons, or preformed Cooper pairs. \cite{Kubo,RMP} The most recent
interest to the system of hard-core bosons comes from the delightful results on
Bose-Einstein (BE) condensed atomic systems.

One of the fundamental hot debated  problems in bosonic physics relates the
evolution of the charge ordered (CO) ground state of 2D hard-core BH model
  with doping away from half-filling. Numerous model studies steadily
confirmed the emergence of "supersolid" phases with simultaneous diagonal and
off-diagonal long range order while Penrose and Onsager \cite{Penrose} were the
first showing as early as 1956 that supersolid phases do not occur.

The most recent quantum Monte-Carlo simulations \cite{Batrouni,Hebert,Schmid}
found two significant features of the 2D Bose-Hubbard model with a
 screened Coulomb repulsion: the absence of supersolid
phase  at half-filling, and a  growing tendency to phase separation (CO-BS)
upon doping away from half-filling.  Moreover, the checkerboard supersolid
phase appears to be unstable thermodynamically. \cite{Batrouni,Hebert,Schmid}
Batrouni and Scalettar  studied quantum phase transitions in the ground state
of the 2D hard-core boson Hubbard Hamiltonian and have shown  numerically that,
contrary to the generally held belief, the most commonly discussed
"checkerboard" supersolid is thermodynamically unstable and phase separates
into solid and superfluid phases. The physics of the CO-BS phase separation in
Bose-Hubbard model is associated with a rapid increase of the energy of a
homogeneous CO state with doping away from half-filling due to a large
"pseudo-spin-flip" energy cost.
Hence, it appears to be energetically more favorable to "extract" extra bosons
(holes) from the CO state and arrange them into finite clusters with a
relatively small number of particles. Such a droplet scenario is believed  to
minimize the long-range Coulomb repulsion.

What is, however, the detailed structure of the CO+BS phase separated state? In
the paper \cite{MBO}  a topological scenario of CO+BS phase separation has been
proposed. The extra bosons/holes doped to a checkerboard CO phase of 2D boson
system are believed to be confined in the ring-shaped domain wall of the
skyrmion-like topological defect which looks like a bubble domain in
antiferromagnet.\cite{AFM} This allows us to propose the mechanism of 2D {\it
topological (CO+BS) phase separation} when the doping of the bare checkerboard
CO phase results in the formation of a multi-center topological defect, which
simplified pseudo-spin pattern looks like  a system of bubble CO domains with
Bose superfluid confined in  charged ring-shaped domain walls.



Such a skyrmion scenario in hc-BH model allows us to draw several important
conclusions.
    The  parent checkerboard CO phase gradually looses its stability under boson/hole doping, while a novel topological self-organized texture is believed to become stable.
 In other words, away from half-filling  one may
anticipate the  nucleation of topological defect in the unconventional form of
the multi-center skyrmion-like object with  ring-shaped Bose superfluid regions
positioned in antiphase domain wall separating the CO core
 and CO outside of the single skyrmion. Specific spatial separation of BS and CO order parameters that avoid each other reflects the competition of kinetic  and potential energy. Such a {\it topological (CO+BS) phase
 separation}
 is believed to provide a minimization of the total energy as compared with
 its uniform supersolid counterpart.

 The most probable possibility is that every domain wall accumulates single boson, or boson
 hole. Then, the number of centers in a multi-center skyrmion nucleated with doping
  has to be equal to the number of doped bosons/holes.

 Apparently, the boson/hole doping is likely to be a driving force
for a nucleation of  a single, or multi-center skyrmion-like self-organized
collective mode,
 which may be (not strictly correctly)
 referred to as
multi-skyrmion system akin in a quantum Hall ferromagnetic state of a
two-dimensional electron gas \cite{Green}. We may characterize an individual
skyrmion by its position (i.e., the center of skyrmionic texture), its size
(i.e., the radius of domain wall), and the orientation of the in-plane
components of pseudo-spin (U(1) degree of freedom). An isolated skyrmion is
described by the inhomogeneous distribution of the CO parameter, or staggered
boson density, charge order parameter characterizing the deviation from the
half-filling, and superfluid order parameter with a modulus and phase.
As regards the charge distribution, the multi-center topological defects has
much in common with a charge-density wave, Wigner crystal or Skyrmion crystal.

It is worth noting that multi-center skyrmions, or multi-skyrmion topological
defects in non-linear $\sigma$-model were regarded by Belavin and Polyakov.
\cite{BP} Interestingly, that such entities may be considered as a system of
non-interacting quasiparticles.

It seems likely that any  doped particle (boson/holes) results in a nucleation
of a new skyrmion, hence its density changes gradually with particle doping.
Therefore, as long as the separation between skyrmionic centers is sufficiently
large so that the inter-skyrmion interaction is negligible, the energy of the
system per particle remains almost constant. This means that the chemical
potential of boson or hole remains unchanged with doping and hence apparently
remains fixed.

\begin{figure*}[t]
\includegraphics[width=18.0cm,angle=0]{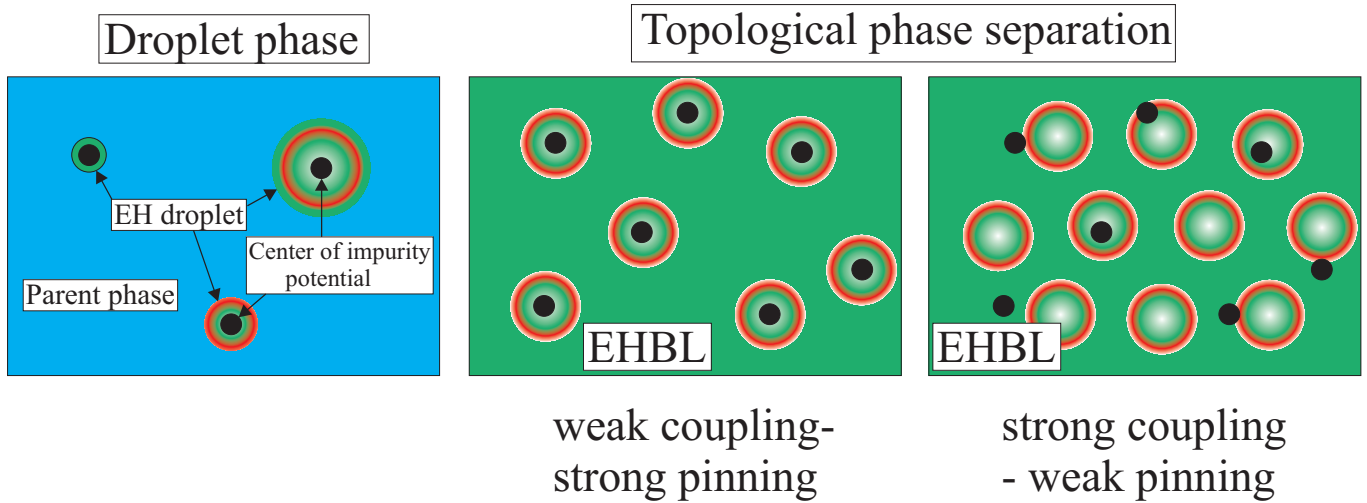}
\caption{Illustrative picture of different phase separation regimes in doped
cuprates} \label{ring}
\end{figure*}

Depending on the doping level we may distinguish three types of topologically
phase-separated  EHBL in doped cuprates: i)"underdoped" EHBL with weakly
coupled skyrmions, when their behavior is ruled mainly by random impurity
potential; ii) "optimally doped" EHBL with strongly coupled skyrmions, forming
skyrmion liquid, or crystal; iii) "overdoped" EHBL with a complex topological
texture that hardly reduces if any to a system of well separated
quasiparticles. The illustrative picture of different phase separation regimes
in doped cuprates is presented in Fig.\ref{ring}.

\section{IR optical response of the doped cuprate.
Static topological phase separation and effective medium theory}

The hole/electron doping into parent cuprate is likely to be a driving force
for a growth of primary electron-hole droplets with a stabilization of  a
single, or multi-center skyrmion-like EHBL collective mode, followed by a first
order phase transition from parent insulating state into unconventional
topological EHBL phase. The appropriate evolution of electron structure should
manifest itself in optical properties. We believe, that in a broad spectral
range the basic features of the optical response of SC oxides can be
consistently accounted for proceeding from the unified physical picture of
their intrinsic electronic inhomogeneity.
 Below we focus on optical response in
infrared. First we address the response of doped cuprate with a static, or
"quenched" phase separation when the crude reasonable model  is that of a
granular composite, whose constituents have the complex dielectric functions
$\eps_i(\w)$, where $i$ labels the species of the grains. What is to be
stressed is that the resulting dielectric function $\eps_{eff}$ of the
composite does not reduce to the simple weighted average of the form
$p\,\eps_1\,+\,(1-p)\,\eps_2$, where $p$ is the volume fraction of its first
component. The correct averaging can be performed within the effective medium
theory and takes into account the specific effects of the resonant scattering
and absorption of light at the interfaces of the grains. The corresponding
surface plasmon resonances in the optical spectra of the composite media are
also referred to as the geometric ones, because their energy is governed by the
shape of the inhomogeneities.

Optical measurements in midinfrared (MIR) region provide an important
information as regards the low-energy excitations  in   cuprates. The nature of
MIR band in insulating cuprates remains to be one of the old mysteries of the
cuprate physics.

 At first glance, overall MIR band in insulating
cuprates may be assigned to different dipole-allowed CT transitions in EH
droplets which certain volume fraction is surely contained in the nominally
pure parent cuprates.
 Generally speaking, the low-frequency optical response of EH droplet
is composed from two main contributions, that of intra-band two-center local
boson transitions and one-center $^1A_{1}-^1E_u$ CT transitions. The former can
form a metallic-like band with the maximal width of the order of 0.1 eV, while
the latter represents the complex optical response of single PJT center which
typical energy falls in the interval of several tenth of eV. However, the EH
droplets are scattered in a highly polarizable surrounding that turns the
calculation of their optical response into a very complicated problem.
  Typical size of the EH droplets seems
to fall in the nanoscopic scale. Hence, when considering the phenomena with a
relevant characteristic length, e.g. infrared optical absorption ($\lambda$ =
0.5 -- 1.5 $\mu$m), doped cuprates can be roughly regarded as a binary granular
medium, composed of two components.

The idea of real-space inhomogeneity is most naturally expressed in terms of an
effective medium theory (EMT), which aims to evaluate some property of a
composite, those of pure components being given. First, we would like to
shortly overview the effective medium theory  \cite{emt} which appears to be a
powerful tool for the quantitative description of the optical response of
inhomogeneous systems. In its simplest form, the EMT equation for effective
dielectric function $\eps_{eff}$ of the two-component inhomogeneous system
reads as follows \cite{emt}:
\begin{equation}\label{ema}
\int \limits_{V} dV p(\eps )
\,\sum_{i=1}^{3}\,\frac{\eps\,-\,\eps_{eff}}{\eps_{eff}\,+\,L_i\,(\eps\,-\,\eps_{eff})}
\,=0,
\end{equation}
where the integration runs over the volume of the sample,
$p(\eps)\,=\,p_1\,\delta(\eps(r)\,-\eps_1)\,+\,p_2\,\delta(\eps(r)\,-\eps_2)$.
We consider the two components of binary composite on equal footing with
$p_{1,2}$ being the volume fractions, $\eps_{1,2}$ and $L_{i}$  the dielectric
functions and the depolarization factors of the grains of its constituents,
respectively.

It is worth to emphasize that in effective medium approximation the volume of
the droplet does not enter the calculations, but is accounted implicitly in the
validity range of the theory, restricting the mean size of the droplet to be
smaller than the wavelength. Hence, within the EMT, the volume fractions
$p_{1,2}$ can only change through the number of the droplets rather than due to
the variation of their sizes. In general, $p_{1,2}$ are determined by
thermodynamical conditions and depend on temperature, pressure, and other
external factors. The approach \cite{PShg}, we employed here to lend more
plausibility to this physically transparent EMT scheme, is to take into account
the natural difference between the core and the surface properties of the EH
droplets using the standard expression for the polarizability of the coated
ellipsoid.\cite{Bilboul}

The spectra of nanoscopically disordered metal-insulator media can display some
specific features due to so-called geometric resonances (surface plasmon, or
Mie resonances)\cite{emt}, that have no counterpart in homogeneous systems.
These arise as a result of resonant behaviour of local field corrections to the
polarizability of the granular composite and are governed to a considerable
extent by the shape of the grains. The frequency of geometric resonance is then
easily obtained as the one at which the polarizability of small particle
diverges. For the case of spherical metallic droplets embedded in the
insulating matrix with a constant dielectric permittivity $\eps_d$ this leads
to the equation:
\begin{equation}
 \eps(\omega)_{m}\,+\,2\,\eps_d\,=\,0,
\end{equation}
whence the resonance frequency is
\begin{equation}
  \omega_r\,=\,\frac{\omega_p}{\sqrt{1\,+\,2\,\varepsilon_d}}, \label{res}
\end{equation}
if simple  Drude's expression
\begin{equation}\label{drude}
\eps\,=\,\eps_0\,-\,\frac{\omega_p^2}{\omega\,(\omega\,+\,\mbox{i}\,\gamma)}
\end{equation}
  with the plasma frequency $\omega_p$ and the damping parameter  $\gamma$ is  assumed
for the metallic permittivity. In the case of arbitrary ellipsoid there are
three different principal values of polarizability and (\ref{res}) generalizes
to
\begin{equation}\label{ell}
\omega_r^i\,=\,\omega_p\,\sqrt{\frac{L_i}{\varepsilon_d\,-\,L_i\,(\varepsilon_d\,-\,1)}},
\ i\,=\,1,\,2,\,3,
\end{equation}
where $L_i$ are three shape-dependent depolarizaton factors. The Drude
approximation is reasonable and is more general that it may seem, since the
form can be regarded as a limiting one of the universal expression for a
dynamical conductivity in terms of memory function  with $\omega_p$ and
$\gamma$ to be effective parameters.

\begin{figure*}[t]
\includegraphics[width=18.0cm,angle=0]{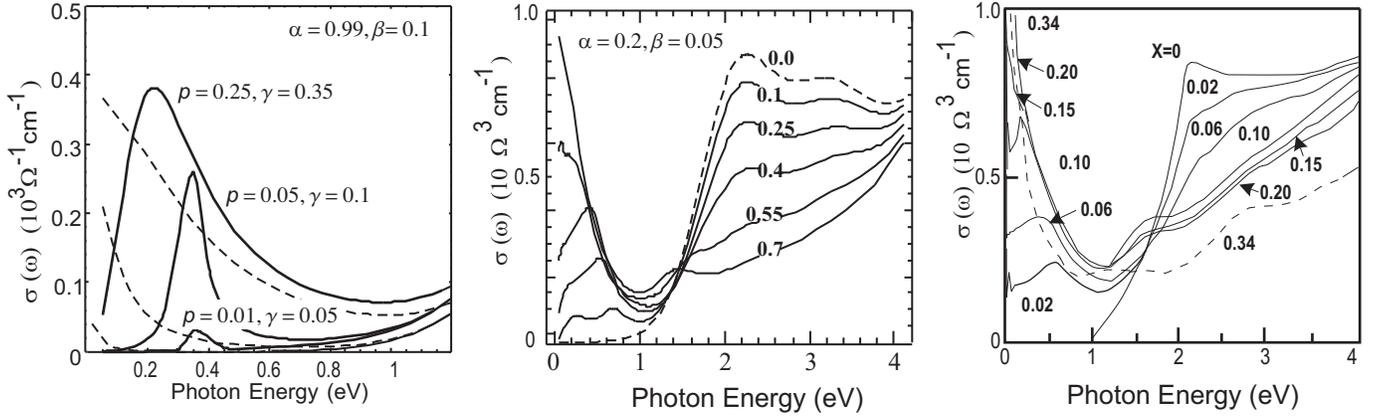}
\caption{Optical conductivity of doped 214 cuprate in frames of the effective
medium theory. Left hand side panel shows  the evolution  of the low-frequency
optical response given small volume fraction of metallic-like  droplets and
varying effective relaxation rate.  Central panel: calculated optical
conductivity for different volume fractions of metallic phase. The numbers
against the curves stand for the relative volume fraction $p$ of the
quasi-metal phase. Right hand side panel: measured optical conductivity of
La$_{2-x}$Sr$_x$CuO$_4$ for different $x$ [\protect\onlinecite{Uchida}]. The numbers against the
curves stand for the doping level.} \label{EMT}
\end{figure*}

An effective medium theory has been applied for the simulation  of the optical
spectra of La$_{2-x}$Sr$_x$CuO$_4$ with $x$ varying in a wide range
\cite{Lumin} (see Fig.\ref{EMT}). Imaginary part of dielectric function of
parent insulating La$_2$CuO$_4$ was fitted to experiment \cite{Uchida} by the
sum of three Gaussians in a broad spectral range to ensure the validity of its
real part, derived via Kramers-Kr\"{o}nig transformation. The phonon bands have
been neglected throughout the calculations. The droplets are assumed to have
ellipsoidal shape with the in-plane semi-axes ratio denoted as $\alpha$ and
that of out-of-plane and major in-plane semi-axes as $\beta$. These geometrical
parameters enter  the theory through the depolarization factors   and were
found to affect strongly the dielectric function of doped cuprate.

Left hand side panel in Fig.\ref{EMT} clearly shows the evolution of the
low-frequency optical response (MIR-band) given small volume fraction of
metallic-like ($\omega _p =1.9$ eV) disc-shaped droplets with varying effective
relaxation rate and volume fraction. For comparison we demonstrate the
predictions of a simple additive two-fluid model $\sigma(\w) = p\,\sigma
_{metal} + (1-p)\,\sigma _{insulator}$ (dotted curves). One may see that the
effect of EH droplets dispersed in a polarizable insulating matrix results in a
crucial rearrangement of the optical response with  the appearance of novel
characteristic features such as Mie resonances. It is worth noting that in
terms of EMT approach  the appearance of metallic-like EH droplets in
insulating matrix leads to the crucial red-shift of the  effective insulating
gap of doped cuprate. In a sense, for lightly doped cuprates one might say
about two characteristic insulating gaps.

 The model in its simplest form
\cite{Lumin} was found able to reproduce all essential features of the
transmittance \cite{Suzuki}, optical conductivity $\sigma(\omega)$, and EELS
spectra \cite{Uchida} for 214 system. It is easy concluded after simple
comparison of the data presented in central and right hand side panels of
Fig.\ref{EMT}.
 Substantial difference in the spectral
and doping dependence of optical conductivity for thin-film \cite{Suzuki} and
bulk samples \cite{Uchida} is easily explained assuming different shape of
metallic and dielectric regions in both materials. New peaks in
$\sigma(\omega)$ and absorption spectra, that emerge in the mid-infrared range
upon doping are attributed to surface plasmon resonances and its interference
with different charge transfer transitions in EH droplets.

Overall, one may conclude that the MIR optical response of insulating cuprates
strongly supports the EH droplet scenario with an intrinsic metallic-like
transport.

\section{IR optical response of the doped cuprate.
Dynamics of  topological defect in a random potential}

The topological structure of EHBL must be considered as being largely dynamic
in nature. Underdoped EHBL can be considered to form a system of weakly coupled
quasiparticles (skyrmions) exposed to the action of a random potential. We
believe, that the model of 2D Bose gas in a random potential offers the most
close physical analogy to our case, and we adopt it as the effective model to
describe the effects of dynamics of  topological defect  in the cuprates.

Below we'll address the approximation of slow topological dynamics with
characteristic energies much lower than the effective insulating gap of doped
cuprate.

\subsection{The memory function technique.}

An adequate formalism for the description of the overdamped charge dynamics in
presence of a strong disorder was elaborated by W. G\"otze and coworkers
\cite{memory,GotzePhil}, who showed, that the optical conductivity $\s(\w)$ at
both sides of Anderson transition can be expressed in a general form:
\begin{equation}
   \s^{\prime} + i\,\s^{\prime\prime} \,=\,i\,\frac{\w_p^2}{\w\,+\,M(\w)}
\end{equation}
where $\w_p$ is the plasma frequency,
$M(\w)=M^{\prime}\,+\,i\,M^{\prime\prime}$ is the memory function. With
$M\,=\,i\,\gamma\,=\,const$ this expression reduces to the simple Drude formula
(\ref{drude0}). Thus, the Drude model  can be understood as a crudest
approximation, when the processes of the current dissipation act equally in the
whole spectral range and do not depend on the frequency.

The real part of the conductivity $\s^{\prime}$, henceforth denoted as $\s$,
may be parameterized in terms of the so called generalized Drude formula
\cite{memory},
\begin{equation}\label{memory1}
   \s(w) = \frac{\Omega_p(\w)^2\,\Gamma(\w)}{\w^2 + \Gamma(\w)^2 },
\end{equation}
where the renormalized plasma frequency $\Omega_p(\w)$ and the relaxation rate
$\Gamma(\w)$ have the form:
\begin{eqnarray}\label{memory2}
   \Omega_p(\w)\,=\,\w_p/\sqrt{\eta(\w)},& \gamma(\w)\,=\,M^{\prime \prime}(\w)/\eta(\w), \\ \eta(\w)=1 + M^{\prime}(\w)/\w.
\end{eqnarray}

The cornerstone result of the approach is based on the mode coupling
self-consistency relation of the memory function and the impurity potential in
the momentum space \cite{GotzePhil}:
\begin{widetext}
\begin{equation}\label{modecoupl}
 M(\w) = \frac{1}{n\,m}\,\int \frac{\Phi_0(k,\, \w + M(\w))}{1 + M(\w)\,\Phi_0(k,\, \w + M(\w))/\chi(k=0, \w)}\, k^2\,\langle |U(q)|^2 \rangle\,d\mathbf{k},
\end{equation}
\end{widetext}
where $n$, $m$ are the concentration and the mass of the carriers, $\Phi_0$ is
the density-density correlation function in absence of the impurities,
expressed through the generalized susceptibility $\kappa(q, z)$:
\begin{equation}\label{fi2}
 \Phi_0(q, z)\,=\,\frac{1}{z}\, \bigl( \kappa(q,z) - \kappa(q)_{z=0} \bigr).
\end{equation}
Within the linear response theory $\kappa(q,z)$ is obtained from the familiar
expression:
\begin{equation}
 \kappa\,=\,-\lim \limits_{\alpha \rightarrow 0}\,\sum \limits_q\,\frac{f(\eps_{k+q})\,-\,f(\eps_q)}{\eps_{k+q}\,-\,\eps_q\,-\w\,-\,i\,\alpha},
\end{equation}
where $\eps_k$ is the energy of the particles and $f$ is their partition
function. Other quantities in Eq. \eq{modecoupl} are the concentration $n$ and
the mass $m$ of the charge carriers.

Now, dwell upon the quantity $\langle |U(q)|^2 \rangle$. This is the average
squared Fourier transform of the random potential $U(\mb{r})$, defined a
superposition of the partial impurity potential $u$, each centered at the sites
$\mb{r}_i$:
\begin{equation}\label{u}
 U(\mb{r})\,=\,\sum \limits_i u(\mb{r}\, - \,\mb{r}_i),
\end{equation}
where the sum is over the impurity  sites. Hence,
\begin{eqnarray}
 \langle |U(q)|^2 \rangle\,&=&\,|u(q)|^2\,s(q) \\
    s(q)\,&=&\,\Biggl\langle \sum \limits_{i,\,j} e^{i\,\mb{q}\,(\mb{r}_i\,-\,\mb{r}_j)}\Biggr\rangle,
\end{eqnarray}
where the brackets denote the averaging over the impurity sites. In what
follows, we shall consider the case of random impurities. Then, the structure
factor $s(q)$ becomes trivial and reduces to $n_i$, the concentration of
impurities, whence
\begin{equation}
\langle |U(q)|^2 \rangle\,=\,n_i\,|u(q)|^2.
\end{equation}

%

In zero temperature limit, the machinery of the correlation functions of free
2D Bose gas is easily implemented because of the particularly simple analytic
form of the generalized susceptibility \cite{hines}:
\begin{equation}\label{fi2}
 \kappa_0(k, z) \,=\, \frac{2 \,n\, \eps_k}{\eps_k^2 - z^2},
\end{equation}
where $\eps_k = k^2/2 m$ is the dispersion of free particles.

To proceed further, it is convenient to introduce the dimensionless variables,
labeled with tilda ($\widetilde{\cdots}$) to distinguish them from their
dimensional counterparts. We have chosen for the units of the mass and the
charge the corresponding quantities for the free electron ($m_e$, $e$). Aiming
to simplify the calculations, we adopt the unit of the wave vector as $q^* = (8
\pi n/a^*)^{1/3}$, where $a^* = \hbar^2 \eps/ m e^2$ is the effective Bohr
radius \cite{gold3dbose}, $\eps$ is the static dielectric constant of the
medium. The unit of energy is then naturally defined as $E^* = \hbar^2 q^{*
2}/2 m_e$. Furthermore, the complex optical conductivity takes the form:
\begin{eqnarray}\label{s0}
  \p{\s} + i\,\s^{\prime\prime} \,=\, \s^* \,\frac{i}{\widetilde{\w}\,+\,\widetilde{M}(\widetilde{\w})}, \\
 \s^*\,=\,\frac{\w_p^2}{E^*}\,=\,\frac{1}{E^*}\frac{n_{3D}\,e^2}{m},
\end{eqnarray}
where $\widetilde{\w}\,=\,\w/E^*,\, \widetilde{M}\,=\,M/E^*$, $n_{3D}$ being
the volume fraction of the droplets. Under moderate doping, we believe this
quantity to be directly related to the doping level $x$:
\begin{equation}
  n_{3D}\,=\,\frac{x}{\Delta^2\,d},
\end{equation}
where $\Delta$ is the lattice constant of the CuO$_2$ plane, $d$ is the
interplane spacing.

Taking into account the numerical values of $m_e$, $e_e$ and $\hbar$, the final
expressions for the adopted units read as follows:
\begin{eqnarray}\label{unitsAll}
  q^*\,&=&\,3.62\cdot 10^8 \,\times \left(\frac{\widetilde{m}\,\widetilde{e}^2\,x}{\eps\,\Delta^2}\right)^{1/3}\,\mbox{cm}^{-1}, \\
  E^*\,&=&\,49.87 \,\times\, \left(\frac{\widetilde{e}^2\,x}{\eps\,\Delta^2\,\sqrt{\widetilde{m}}}
\right)^{2/3}\,\mbox{eV}, \\
 \s^*\,&=&\, 4.0\cdot
10^3\,\times\left(\frac{\widetilde{e}^2\,\eps^2\,x}{\Delta^2\,d^3\,\widetilde{m}^2}
\right)^{1/3}\,\Omega^{-1}\,\mbox{cm}^{-1},
\end{eqnarray}
where we made use of the relation 1 eV
$\simeq\,1.824\cdot10^3\;\Omega^{-1}\,\mbox{cm}^{-1}$. Let us estimate their
possible numerical values. Consider the situation, when the EHBL phase is
composed of the heavy Bose particles carrying double elementary charge,
$\widetilde{m} = 10,\, \widetilde{e} = 2$. Given the parameters of \LSCO unit
cell, $\Delta \simeq 4$ \AA, $d \simeq$ 6 \AA, and the bare dielectric constant
$\eps = 40$, we obtain for slight doping $x = 0.01$ the following units of
energy and conductivity: $E^*\,\simeq\,294.5\,\mbox{cm}^{-1}$, $\s^* \,\simeq\,
0.23 \cdot 10^3\,\Omega^{-1}\,\mbox{cm}^{-1}$.

Making use of these notations, it is straightforward to write down the explicit
expressions of the correlation functions of the theory:
\begin{eqnarray}
 \kappa(q, z) = (2 n/E^*)\,\widetilde{\kappa}(\widetilde{q}, \widetilde{z}) \\
 \Phi(q, z) = (2 n/E^{*\,2})\,\widetilde{\Phi}(\widetilde{q}, \widetilde{z}) \\
             (z\,=\,E^*\,\widetilde{z},\;q\,=\,q^*\,\widetilde{q}) \nonumber,
\end{eqnarray}
where the frequency and wave vector dependencies are contained in
dimensionless expressions:
\begin{equation}
 \widetilde{\kappa}(\widetilde{q}, \widetilde{z})_{RPA}\, = \,\frac{\widetilde{q}^2}{\widetilde{q}^4 +
\widetilde{q} - \widetilde{\w}^2},
\end{equation}
\begin{equation}\label{corrfunc}
\widetilde{\Phi}(\widetilde{q}, \widetilde{z}+\widetilde{M})\, =
\,\frac{\widetilde{q}\,(\widetilde{z} + \widetilde{M})}{(1 +
\widetilde{q}^3)(\widetilde{q} + \widetilde{q}^4 - (\widetilde{z} +
\widetilde{M}) \widetilde{z})},
\end{equation}
and
\begin{equation}
 \kappa(k, z)_{RPA} = \frac{\kappa(k, z)}{1 + 2 \pi e^2\kappa(k, z)/k}
\end{equation}
takes into account the 2D  Coulomb interaction of quasiparticles in a random
phase approximation.

It is also convenient to isolate the dimensionless $\widetilde{q}$- dependence
in the impurity potential:
\begin{equation}\label{uf}
 u(q)\,=\,U_0\,f(\widetilde{q}),
\end{equation}
where $U_0$ has the dimension of energy and $f$ specifies the shape of the
potential.
\begin{figure*}
 \centering\includegraphics[width=6.0in,height=3.5in]{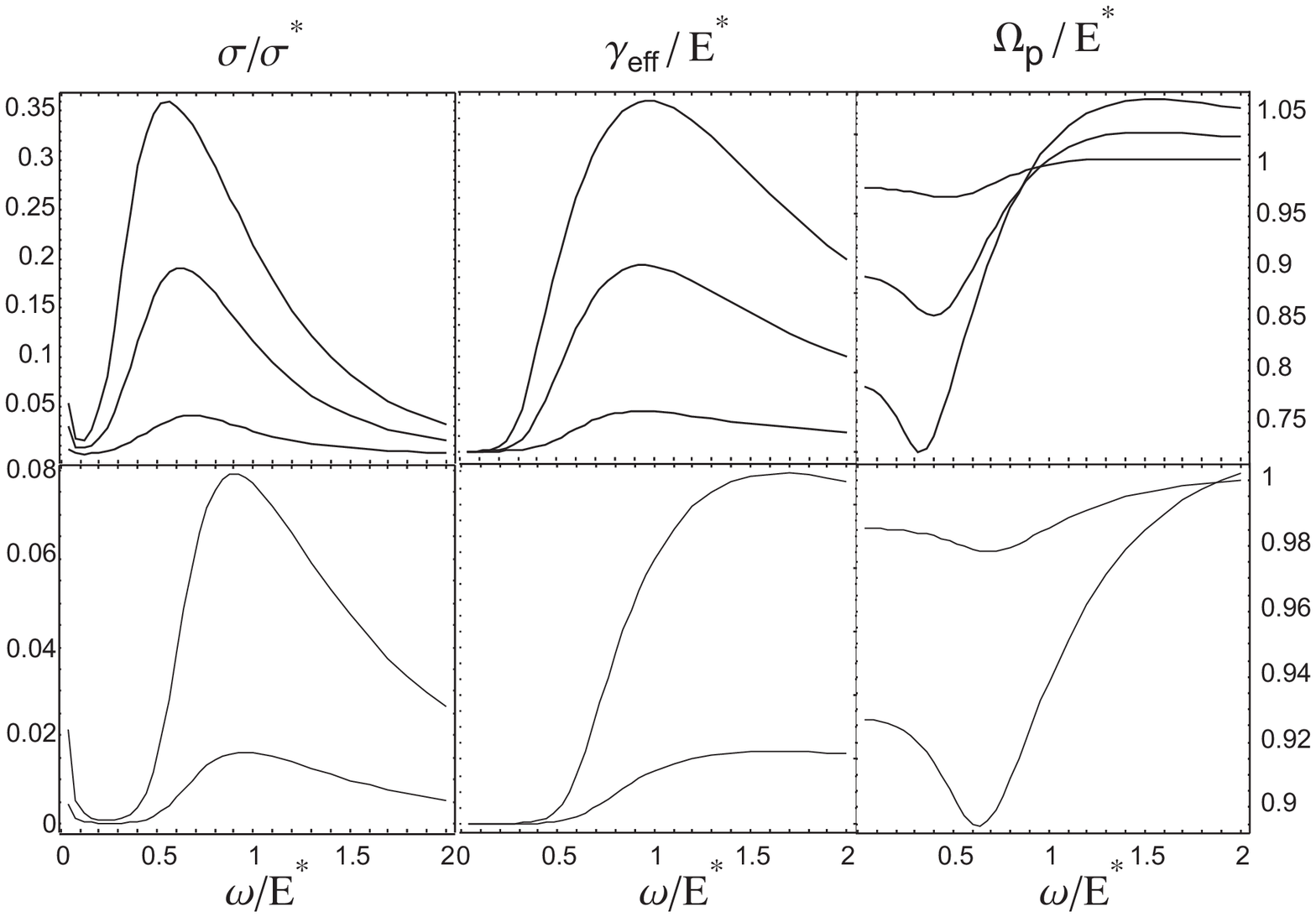}
 \caption{Dimensionless optical conductivity and the parameters of the generalized
 Drude expression \eq{memory2} for different strength of disorder (upper curves
 correspond to stronger disorder). Upper row: Coulomb impurities. Lower row: Debay
 impurities.}
\label{dless}
\end{figure*}
Finally, the dimensionless analogue of Eq. \eq{modecoupl} has the form:
\begin{equation}\label{mdless}
 \widetilde{M} = \Upsilon\, \int \limits_0^{\infty} d\widetilde{q}\; \widetilde{q}^3\,\widetilde{\Phi}(\widetilde{q}, \widetilde{z}+\widetilde{M}) \,|f(\widetilde{q})|^2,
\end{equation}
Thus, the dimensionless memory function is governed by the only parameter,
which is defined as: $\Upsilon = 4 \widetilde{U}_0^2\,n_i/n$, where
$\widetilde{U}_0$, measured in units of $E^*$, is the strength of the potential
(the depth of the potential well), $n_i$ and $n$ are the concentration of the
impurities and the carriers, respectively.

\subsection{Model simulations}

\subsubsection{Point charges in random potential.}

The Eqs. (\ref{corrfunc} - \ref{mdless}) provide the practical scheme to
calculate the optical response of the system in presence of the random
potential. We begin by considering the case of the usual 2D charged Bose gas in
random potential.

Two especially interesting cases of the impurity potentials have been examined.
The first one is the case of Coulomb impurities, located at a certain distance
$h$ over the CuO$_2$ planes: $u(r) = U_0\,(R^2 + h^2)^{-1/2}$, where $R$ is the
in-plane projection of $r$. The 2D Fourier transform of the potential is $u(q)
= U_0\,\exp(- q\,h)/q$. The second case we considered here corresponds to the
Debay impurities, $u(r) = U_0 \exp(-r/r0)/r$, where $r_0$ is the screening
length. The Fourier transform is $u(q) = U_0\,r_0\,(1 + q^2\,r_0^2)^{-1/2}$.

The overall view of dimensionless spectral dependencies of optical
conductivity, effective relaxation rate $\Gamma(\w)$ and the effective plasma
frequency $\Omega_p(\w)$ (\ref{memory1}, \ref{memory2}) is shown in Fig.
\ref{dless}. It can be seen, that the conductivity spectrum is characterized by
a broad asymmetric feature due to plasma oscillations of quasiparticles in the
potential wells at the plasma frequency, and is not very sensitive to the
choice of the profile of the wells, specified by the impurity potential. The
stronger is the localization, the sharper is the plasma peak. Because of the
total spectral weight conservation, $\int_0^{\infty} \s(\w)\,d\w\,=\,const$,
this implies the transfer of the intensity from the coherent response of the
condensate $\propto\,\delta(\w)$ to the plasma modes.

To be noted is a strong frequency dependence of effective relaxation rate
$\Gamma(\w)$ with nearly constant slope in a wide range, reminiscent of the
marginal Fermi liquid behaviour  of the cuprates.

The comparison with the previously studied 3D Bose gas \cite{gold3dbose}, 3D
and 2D Fermi gas \cite{GotzePhil, GoldFermi} makes it possible to conclude,
that the response of localized carriers always remains essentially the same.
Still more nontrivial is, however, the resemblance of these spectra with those,
obtained for pinned charge density waves \cite{fukuyama} and Wigner crystals
\cite{giamarchi}. It seems, that there is a certain dualism between the
collective modes of localized charges and pinned phase degrees of freedom.


\subsubsection{Skyrmion-like quasiparticles in a random potential.}

As far as we treat the real-space electronic inhomogeneity in oxides within the
effective model of 2D Bose gas, the basic results of previous section hold also
for this case.

However, the topological texture of EHBL phase is not the system of
structureless point quasiparticles, but rather skyrmion-like entities with a
complicated internal structure. Hence, the explicit form of the effective
random potential does not remain as simple as what follows from the Coulomb
law. Nevertheless,  among the possible interactions of the charged impurity
with the inhomogeneity  the direct electrostatic one is of prime importance.

It is possible to model the charge distribution within the quasiparticle in a
direct manner, assigning the charge density $\rho_i$ to its point (Fig.
\ref{inh}), if the discrete nature of the quasiparticle in the CuO$_2$ is taken
into account. The interaction of an impurity with the quasiparticle is
obtained, summing over "internal" coordinates of the quasiparticle (see Eq.
\eq{uf}):
\begin{equation}
 u(\mb{r})\,=\,\sum \limits_i\,\rho_i\,U_0\,f(\mb{r} - \mb{r}_i),
\end{equation}
whence the Fourier transform is straightforwardly obtained:
\begin{equation}
 u(\mathbf{k})\,=\,U_0\,f(k)\,F(\mathbf{k}),
\end{equation}
where
\begin{equation}
 F(\mathbf{k})\,=\,\sum \limits_l \, \rho_l\,\exp(i\,\mathbf{k}\cdot\mathbf{r}_l)
\end{equation}
is the form-factor of the droplet and $f(k) = e^{-k\,h}/k$ for Coulomb
impurities, randomly spread at a distance $h$ over the CuO$_2$ planes.

For the applications, we have to calculate the expression $\langle |F(k)|^2
\rangle$, averaged  over the orientations of the quasiparticles. To this end,
the additional physical assumptions about the arrangement of the quasiparticlet
within the CuO$_2$ plane is required. Since there are {\it a priori} no reasons
to expect some ordering of the quasiparticle, we consider below the case of
they random orientations. In two dimensions, we obtain:
\begin{widetext}
\begin{equation}\label{FF}
  \langle |F(k)|^2 \rangle\,=\,\frac{1}{2\,\pi} \int \limits_0^{2\,\pi}\, \sum \limits_{i\,j} \, \rho_i\,\rho_j\, e^{i\,k\,(r_i - r_j)\cos \,\varphi}\,d\varphi\,=\, \sum \limits_{i\,j} \rho_i\,\rho_j\,J_0(k\,r_{ij}),
\end{equation}
\end{widetext}
where $r_{ij} = |\mathbf{r}_i - \mathbf{r}_j|$ is the distance between two
arbitrary internal points of the quasiparticle, $J_0$ is the Bessel function of
the first kind.

The form-factor of the quasiparticle enables us to investigate a number of the
charge configurations of particular interest, such as the charged linear
stripes etc., of course with reserve, that  only the details of radial charge
distribution enter the average form-factor.


-----------------------------------
-----------------------------------

It is therefore instructive to consider first the simple case, when the charge
distribution in a quasiparticle has the form of an uniform ring with inner
radius $R_1$ and outer radius $R_2$. We denote their ratio as $\alpha = R1/R2$.
Changing $\alpha$ continuously from 0 to 1, we can model both the disk, the
thin hook and the intermediate ring configurations. The continuous counterpart
of Eq. \eq{FF} reads:
\begin{equation}
 \langle |F(k)|^2 \rangle(q)\,=\, \Biggl( 2\,\pi\,\int \limits_{R_1}^{R_2}\,r\,\rho(r)\,J_0(q\,r)\,dr \Biggr)^2,
\end{equation}
where $\rho\,=\,Q / \pi\,(R_2^2 - R_1^2)$ -- is the (constant) charge density,
$Q$ is the full charge of the ring.

--------------------------------------------------------
\begin{figure*}[!]
\begin{minipage}[b]{0.5\linewidth}
\includegraphics[width=0.95\linewidth,angle=0]{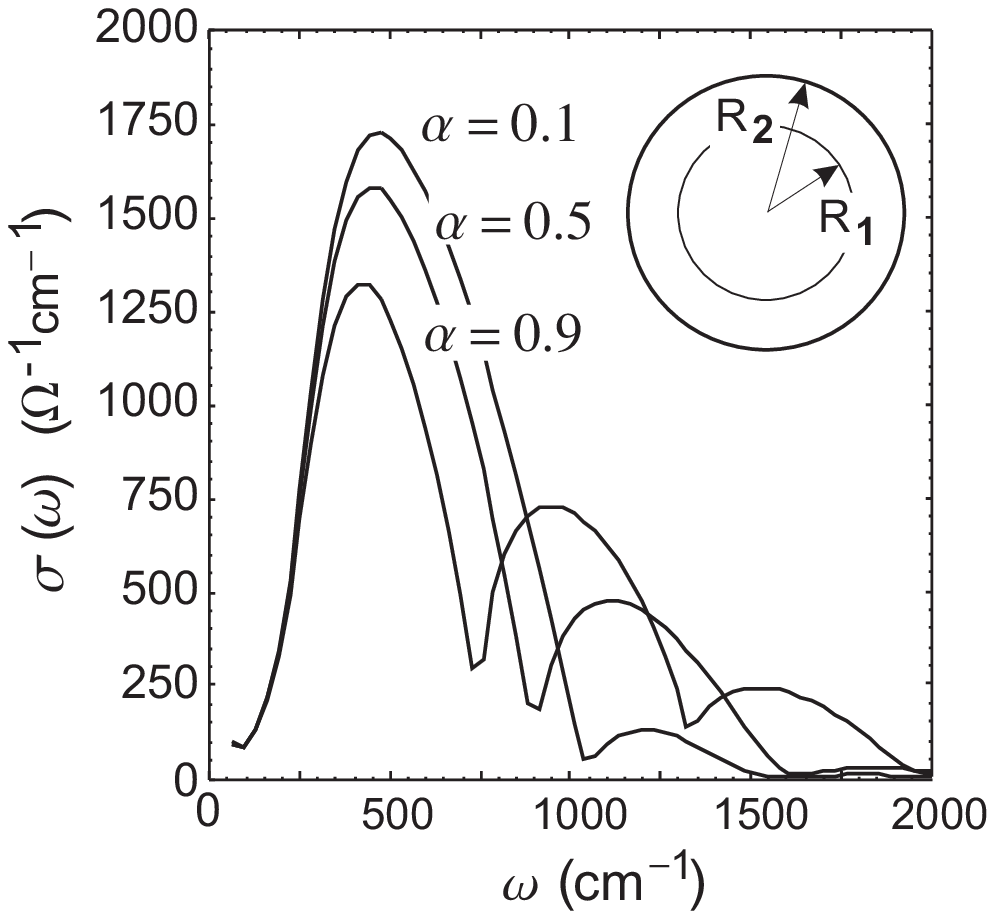}
 \end{minipage}\hfill
 \begin{minipage}[b]{0.48\linewidth}
\includegraphics[width=\linewidth,angle=0]{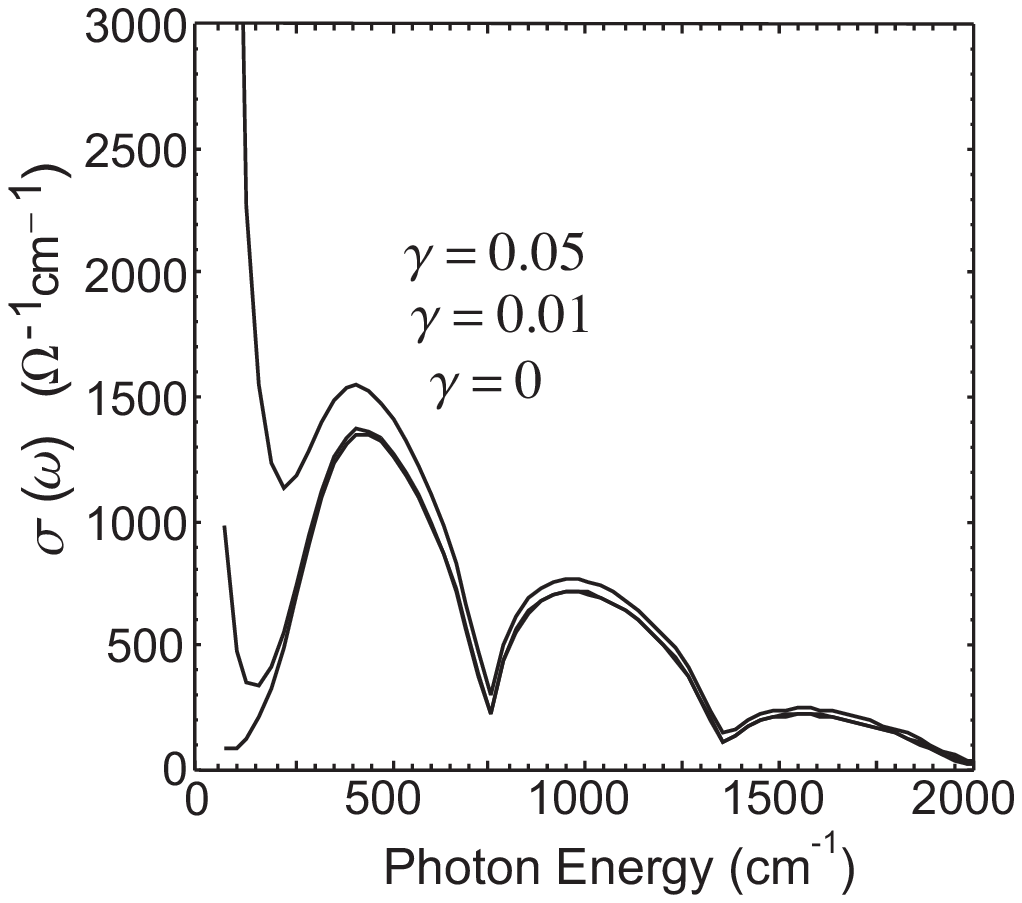}
 \end{minipage}
\caption{The optical conductivity of the skyrmion-like quasiparticles with the
ring-shaped charge distribution in a random potential.  Left panel illustrates
the role of different inner-to-outer radius ratio denoted by $\alpha$. Right
panel  illustrates the role of different values of the bare relaxation
parameter $\gamma_0$ = 0.0, 0.01, and 0.05 eV given $\alpha = 0.9$}
\label{rings}
\end{figure*}
--------------------------------------------------------

A series of model optical conductivity spectra, calculated for the ring
geometry of charge distribution in skyrmion-like quasiparticle, is depicted at
Fig. \ref{rings}. It shows the  spectra of the bosonic quasiparticles (mass $m
= 2 m_e$, charge $2\,e$) distributed within the CuO$_2$ plane with the
concentration $n_{2D} = 3.5\,10^{-3}$ \AA$^{-2}$ and trapped by the  potential
induced by Coulomb impurities randomly located at the height $\sim 3 \AA$ above
CuO$_2$ plane with the mean wells depth $\sim\,1500$ cm$^{-1}$.  The value of
the static dielectric constant $\eps = 40$ was used in the calculation. The
outer radius of rings was fixed to be 30 $\AA$.

As the aperture of the ring increases, its outer radius kept constant, the
intensity of the plasma peak progressively drops because the charge gradually
moves away from the impurity. In other words, the larger is the quasiparticle,
the weaker it is trapped by the potential. However, the increase of the
quasiparticle extent does not simply reduce to its gradual delocalization. It
can be seen, that spectrum of the plasma peak of the ring decays into the
oscillating structure, that becomes more pronounced as the ring gets thinner.
This means, that the inhomogeneous charge distribution in the quasiparticle
leads to the complicated profile of the collective mode spectrum with a fine
resonant structure.
 To get insight into the nature of these oscillations, we
remember, that the frequency dependence of the memory function $M(\w)$ is
obtained, according to Eq. \eq{mdless}, as a result of the integration of a
complicated density fluctuation correlation functions over the momentum space.
Taking account of the form-factor of quasiparticle can be interpreted as the
evaluation of the memory function using the same integrand, as for the case of
structureless point particles, now with a highly nontrivial $k$-dependent
plasmon density of states, that is tantamount to the modification of the
dispersion relation of the quasiparticles as distinct from the case of the free
Bose gas.

The  result obtained is to be compared with the microwave optical conductivity
of lightly doped \LSCO \cite{hor} ($x = 0.015$). The multi-peak structure of
calculated spectra falls in nearly the same spectral range, as that, observed
in experiment. The authors of the paper [\onlinecite{hor}] regard the
resonances in optical conductivity to be the manifestation of the pinned Wigner
crystal response of the carriers in CuO$_2$ planes. We believe, that our
results could provide another (quantitative) interpretation of these
experimental findings in terms of the plasma oscillations of the quasiparticles
phase, trapped in the roughness of a complicated non-uniform potential
landscape.


The memory function approach  considered here enables us to model the optical
response of the quasiparticles due to their collective plasma modes. Of course,
there are also another channels of the dissipation in a phase-separated
systems, related to the excitations of other degrees of freedom, that are
beyond the scope of the present analysis. We may try to get idea of the role of
this additional absorption mechanisms phenomenologically, inserting an
additional relaxation term in the denominator of the generalized Drude formula.
This trick reflects the fact, that different mechanisms of current relaxation
act nearly independently. Having in mind to minimize the number of model
assumptions, we consider the following trial formula
\begin{equation}\label{effM2}
 \s(\w)\,=\,\mbox{Re}\,\frac{\Omega_p^2}{\w\,+\,M(\w)\,+\,\mbox{i}\,\gamma_0},
\end{equation}
where $\gamma_0 = const$. Note that the trivial memory function
$M\,=\,\mbox{i}\gamma_0$ corresponds to the conventional Drude formula.

The right hand side panel in Fig. \ref{rings} illustrates the effect of a
gradual increase of the bare relaxation parameter $\gamma_0$ on the optical
conductivity of the "ring-shaped" quasiparticles. In absence of $\gamma_0$ the
optical conductivity displays the broad plasma feature and pseudogap depression
of the spectrum at lower frequencies. Turning on $\gamma_0$ leads to the
development of the Drude like tail within the gap.

It is interesting to note that such a spectral behavior is actually observed
for  different cuprates (see e.g. Ref.\onlinecite{startseva}) with varying the
temperature. It may be seen, that the inclusion of additional dissipation
channel allows to simulate the temperature effect.

\section{Conclusion}

 We consider the cuprates  to be unconventional systems which are
unstable with regard to a self-trapping  of the low-energy charge transfer
excitons with a nucleation of EH droplets being actually the system of coupled
electron CuO$_{4}^{7-}$ and hole CuO$_{4}^{5-}$ centers having been glued in
lattice due to  strong electron-lattice polarization effects. The hole/electron
doping into parent cuprate is likely to be a driving force for a growth of
primary EH droplets with a gradual stabilization of  a single, or multi-center
skyrmion-like EH Bose liquid collective mode, followed by a first order phase
transition from parent insulating state into unconventional topological EH Bose
liquid phase. The latter is believed to be described as a system of interacting
skyrmion-like entities which concentration is specified by the doping.
Nanoscopic electron inhomogeneity is believed to be inherent property of doped
cuprates throughout the phase diagram beginning from EH droplets in insulating
parent system and ending by a topological phase separation in EHBL phase.

We have examined the effects of electron inhomogeneity accompanying such a
phase separation on IR optical conductivity. A simple model of metal-insulator
composite and effective medium theory has been used to describe the static
phase separation effects. Specific effects of spectral weight red-shift with
the appearance of low-energy effective insulating gap, the seeming "quenching"
of effective plasmon frequency, and the emergence of MIR bands have been
assigned to the effects of phase separation.

The low-frequency dynamics of topological EHBL phase in a random potential  in
underdoped regime has been discussed in a quasiparticle approximation in frames
of the memory function formalism. The appropriate low-frequency optical
conductivity appears to have anomalous many-peak "non-Drude" spectral
dependence that resembles that of observed experimentally.

The effects of static and dynamic nanoscopic phase separation are believed to
describe the main peculiarities of the optical response of doped cuprates in a
wide spectral range.

\begin{acknowledgments}

Authors acknowledge the
 support by  SMWK Grant, INTAS Grant No. 01-0654, CRDF Grant No. REC-005,
RME Grant No. E02-3.4-392 and No. UR.01.01.042, RFBR Grant No. 01-02-96404.

\end{acknowledgments}

\end{document}